\documentclass[acmsmall]{acmart}

\AtBeginDocument{%
  }

\setcopyright{none}
\copyrightyear{2025}
\acmYear{2025}
\acmDOI{XXXXXXX.XXXXXXX}

\usepackage{xspace}
\usepackage{url}

\newtheorem{definition}{Definition}

\newcommand{\ie}{\emph{i.e.,}\xspace}
\newcommand{\eg}{\emph{e.g.,}\xspace}

\begin{document}

\title{A Task-Centric Perspective on Recommendation Systems}

\author{Aixin Sun}
\email{axsun@ntu.edu.sg}
\orcid{0000-0003-0764-4258}
\affiliation{%
  \institution{Nanyang Technological University}
  \city{Singapore}
  \country{Singapore}
}

\begin{abstract}  
Many studies in recommender systems (RecSys) adopt a general problem definition, \ie to recommend preferred items to users based on past interactions. Such abstraction often lacks the domain-specific nuances necessary for practical deployment. However, models are frequently evaluated using datasets collected from online recommender platforms, which inherently reflect domain or task specificities.
In this paper, we analyze RecSys task formulations, emphasizing key components such as input-output structures, temporal dynamics, and candidate item selection. All these factors directly impact offline evaluation. We further examine the complexities of user-item interactions, including decision-making costs, multi-step engagements, and unobservable interactions, which may influence model design. Additionally, we explore the balance between task specificity and model generalizability, highlighting how well-defined task formulations serve as the foundation for robust evaluation and effective solution development. 
By clarifying task definitions and their implications, this work provides a structured perspective on RecSys research. The goal is to help researchers better navigate the field, particularly in understanding specificities of the RecSys tasks and ensuring fair and meaningful evaluations.
\end{abstract}  
%%
%% The code below is generated by the tool at http://dl.acm.org/ccs.cfm.
%% Please copy and paste the code instead of the example below.
%%


%%
%% Keywords. The author(s) should pick words that accurately describe
%% the work being presented. Separate the keywords with commas.


%\received{20 February 2007}
%\received[revised]{12 March 2009}
%\received[accepted]{5 June 2009}

\settopmatter{printfolios=true}

\maketitle

%====================
\section{Introduction}
\label{sec:intro}
%====================

In many recommender systems (RecSys) studies, solutions are proposed for a commonly adopted problem: given a set of users, items, and their interactions, the goal is to recommend items that align with users' interests or preferences. While this general problem definition captures common patterns across recommendation scenarios, its abstraction overlooks critical details needed for practical RecSys applications. Furthermore, discussions on task definition often lack clarity, particularly regarding its scope and practical implications.
At the same time, RecSys research is closely tied to real-world applications, where models are primarily evaluated using datasets obtained from operational recommender platforms. The mismatch between abstract problem definitions and domain-specific evaluations leads to inconsistent settings and findings across experiments~\cite{EvaluationSurvey22,McElfreshKV0W22}.

We begin  with a review of several highly cited works in RecSys, emphasizing task formulations. Interestingly, key factors in RecSys were well defined and discussed decades ago. Yet, the community still disagrees on the choice of baselines and datasets~\cite{RecBaselines23,DatasetRecsys24}. We then provide a detailed examination of task definition, focusing primarily on the input and output of a mapping function, \ie the recommender. We discuss the missing elements in task formulation: time and the selection of candidate items for recommendation, and their impact on offline evaluation. Lastly, we review  the life cycle of user-item interactions, with a focus on the cost incurred from recommendations made to the feedback of the user-item interactions.
Based on the task definition and proposed framework, we outline key perspectives and actionable directions for future work.

%================================
\section{A Historical Review of Task Formulation}
\label{sec:hisReview}
%================================

We begin with a few widely cited RecSys papers~\cite{HerlockerKTR04,KorenBV09,RicciRS11Chp,TkdeSurvey05}. As foundational works in this field, these papers have influenced many researchers, and the tasks they define have likely shaped numerous follow-up studies. In an influential survey paper, \citet{TkdeSurvey05} formally define the \textit{recommendation problem} as follows: 
\begin{definition}[Recommendation Problem]
\label{def:orgDef}
Let $U$ be the set of all users and let $I$
be the set of all possible items that can be recommended. Let $r$ be a utility function that measures the usefulness of item $i$ to user $u$, \ie $r\in U\times I\rightarrow R$, where $R$ is a totally ordered set (\ie nonnegative integers or real numbers within a certain range). Then, for each user $u\in U$, we want to choose such item $i'\in I$ that maximizes the user's utility:
\begin{equation}\label{eqn:general}
    \forall u\in U, i'_u=\arg \max_{i\in I} r(u,i)
\end{equation} 
\end{definition}
The authors further note that ``the utility of an item is usually represented by a rating, which indicates how a particular user liked a particular item,'' a concept commonly known as \textit{explicit feedback}. In subsequent RecSys studies, \textit{implicit feedback} has become far more prevalent~\cite{ImplicitFeeback09}. As a result, the utility of an item to a user is often inferred from binary feedback, whether the user has interacted with the item. Notably, the difference between explicit and implicit feedback primarily affects data modeling and the loss function design when learning $r$, while the core recommendation problem remains unchanged.

\citet{KorenBV09} compare the two primary approaches in RecSys: content filtering and collaborative filtering. Content filtering builds user and item profiles based on their characteristics, allowing the system to match users with relevant items. In contrast, collaborative filtering relies exclusively on past user-item interactions. Although solutions such as content-based, collaborative, and hybrid filtering are independent of problem definitions, the widespread adoption of collaborative filtering in recent research leads us to assume that \textit{user-item interactions remain a key input for typical recommender systems}. Regarding user preferences,  \citet{RicciRS11Chp}  highlight that these can also be inferred from their actions, such as navigating to a specific product page. In this context, user feedback can take multiple forms: \textit{explicit} feedback, such as ratings; \textit{implicit} feedback, derived from user-item interactions; and \textit{inferred preferences} based on observed user behavior.

\citet{HerlockerKTR04} provide a thorough discussion on user tasks for recommender systems. Their discussion focuses on end-user tasks (\ie not marketers or other system stakeholders)\footnote{We refer readers to~\cite{EvaluationSurvey22} for more detailed discussion on other users in a recommender system like item provider, platform provider, and other stakeholders.}, which aligns well with the RecSys tasks to be discussed in this paper. The key user task is to find good items, such as providing users with a ranked list of recommended items. The authors highlight that ``there are likely to be \textit{many specializations of the tasks within each domain},'' and the domain-specific characteristics are reflected in the properties of the datasets. 

%====================
\section{A Closer Look at the Task Definition}
\label{sec:task}
%====================

For clarity, we rewrite Definition~\ref{def:orgDef} by specifying a recommender as a mapping function.
\begin{definition}[Recommendation]
\label{def:basic}
Let $U$ be a set of users and $I$ be a set of items. A recommender  aims to produce a ranked list of items for a user $u$, based on user-item interactions $U \times I$. 
\begin{equation}\label{eqn:basic}
\langle u, U\times I, I \rangle \rightarrow R^u
\end{equation}
\end{definition}
In this rewritten form, a recommender's input consists of three components: 
(i) the user $u$ for whom recommendations are to be made,  
(ii) the set of user-item interactions $U \times I$, which includes  existing interactions made by users and items, and
(iii) the set of available items $I$ from which recommendations are being made.  The system outputs a ranked list of recommended items for $u$, denoted by $R^u$. This definition is generic to cover all attributes or side information of users or items, because all such information can be easily derived from user IDs and item IDs. 

%====================
\subsection{The Missing Input: Time}
%====================

In practical scenarios, whether recommending a product for purchase or a song to listen to, recommendations made at a time point $t$ should be based on all information available at $t$. With that, we rewrite the mapping function in Definition~\ref{def:basic} to consider the time dimension. We also make an assumption that each interaction between a user $u$ and item $i$ is associated with a time stamp $t_x$ when the interaction occurred, denoted by $(u, i, t_x)\in U\times I$.
\begin{equation}\label{eqn:time}
\langle u, t, (U\times I)_{\leq t}, I_{\leq t} \rangle \rightarrow R_t^u  \end{equation}
Here, we introduce a time point $t$, indicating the time a recommendation is to be made for user $u$. We use $(U \times I)_{\leq t}$ to represent all user-item interactions that occurred before time $t$, a simplified form of \{$(u, i, t_x)\in U \times I| t_x\leq t$\}. The candidate items available for recommendation are those that were registered in the system and accessible as at $t$, denoted by $I_{\leq t}$.

\begin{figure}
    \centering
    %\fbox{
    \includegraphics[clip, trim=0.7cm 13cm 13.8cm 3.2cm, width=0.45\linewidth]{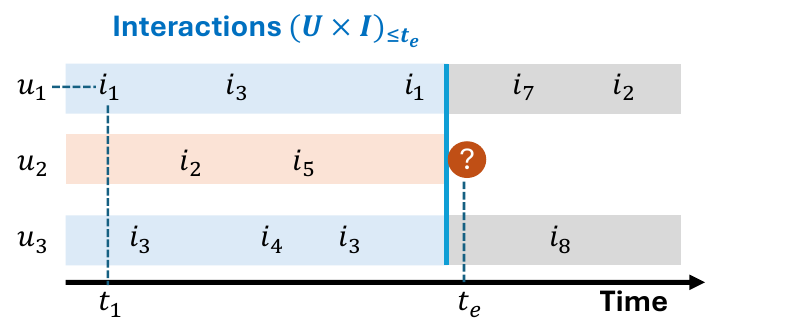}
    %}
    \caption{Recommendations are to be made for user $u_2$ at time point $t_e$. A model is expected to learn from interactions occurred before $t_e$ \ie $(U\times I)_{\leq t_e}$ and recommend items available at $t_e$, denoted by $I_{\leq t_e}$.}
    \label{fig:taskTime}
    \Description{An illustration of the temporal relationship between a test instance and training instances}
\end{figure}

While this reformulation may seem trivial and does not impact real-world or online RecSys implementations, strictly adhering to this task definition poses significant challenges for offline evaluations~\cite{FreshSun23}.  In offline evaluations, user-item interactions in a dataset are partitioned into training and test sets. The former is used to train a recommendation model, while the latter is used to assess its performance. 
Figure~\ref{fig:taskTime} provides an illustration where the items interacted with by each user are plotted to the right of the user along a global timeline. For example, user $u_1$ interacted with item $i_1$ at time point $t_1$.  If we use the user-based leave-last-one-out data split to evaluate our model's accuracy for user $u_2$, the last interaction of $u_2$ at time $t_e$ is masked as the test instance. The recommender can learn from all user-item interactions that occurred before $t_e$, \ie $(U \times I)_{\leq t_e}$. The items available for recommendation to $u_2$ at $t_e$ are those that have been interacted with by any user before $t_e$: $I_{\leq t_e}=\{t_1, t_2, t_3, t_4, t_5\}$.  Notably, items $i_7$ and $i_8$ received their first interaction after $t_e$, the system has no interaction data for these items at $t_e$, and they should not be recommended to $u_2$.\footnote{Note that our discussion assumes the recommendation model is based on collaborative filtering, to avoid the possible misunderstanding that items $i_7$ and $i_8$ could be recommended based on profile matching.} 

Note that the user-based leave-last-one-out data split results in each user having their own $(U \times I)_{\leq t}$ and $I_{\leq t}$, since not all users have their last interaction at the same time. Without enforcing a global timeline or an absolute time point, many commonly used data partitioning schemes, like random splitting or user-based leave-last-one-out, can lead to data leakage. Empirically, we show that the impact of data leakage on recommendation models is unpredictable~\cite{DataLeakage23}. Hence, comparing the performance of recommendation models under data leakage in offline evaluation is both impractical and meaningless.
Recently, \citet{Le25Dont} further show that, for sequential recommenders, eliminating data leakage leads to a 21.7 - 73.4\% drop in sampled nDCG@10 compared to the commonly adopted setting that includes data leakage. They also observe that the impact of data leakage on model rankings is unpredictable, consistent with our findings in~\cite{DataLeakage23}.

Indeed, some recent studies have realized the importance of this issue, and adopted splits based on absolute temporal cutoffs. To examine how prevalent this practice is, we reviewed the 49 long papers published at the ACM RecSys 2025 conference (September 2025). We found that 16 papers adopted random splits, entirely ignoring the temporal factor. Among the remaining papers that report experimental settings, 7 used relative temporal splits, which also lead to data leakage.

The incorporation of the temporal factor $t$ into the task formulation also directly affects the implementation of certain baseline models. For example, the widely used popularity baseline (which simply counts item occurrences in the training data) does not accurately reflect real-world item popularity. In practice, item popularity is typically defined with respect to a reference time point and within a specific time window, \eg the top-selling books from the past week or month. By explicitly introducing a time point $t$ into the problem definition, item popularity can be computed relative to that specific time point. Interestingly, the simple temporally aware popularity model, DecayPop~\cite{PopularitySigir20}, has been shown to be the most effective ranking method on the Yambda-5B dataset for the music recommendation `like' scenario, outperforming many more complex models~\cite{Yambda5b25}. This finding calls for a careful re-evaluation of prior studies that overlooked the temporal dimension.

%====================
\subsection{The Missing Constraints on Candidate Items}
%====================
\citet{HerlockerKTR04} discuss recommendation tasks from the perspective of novelty and quality, assuming that users generally expect recommended items to be new and previously unconsumed. In some scenarios, users may prefer to interact with items they are already familiar with and confident in~\cite{anderson2014dynamics}.
Groceries shopping~\cite{ariannezhad2022recanet,katz2022learning},  e-commerce~\cite{wang2019modeling,quan2023enhancing}, food delivery ordering~\cite{Jiayu24}, and song listening~\cite{tsukuda2020explainable,reiter2021predicting} are a few examples where a user may prefer to have earlier consumed items to be recommended for easy selection.

Let $I^u_t$ be the items that $u$ has interacted with in the past at time point $t$. We use $I_t^{\bar{u}}$ to denote the remaining items available at $t$ that are new to $u$.  Equation~\ref{eqn:time} can be divided into two formulations for repeated consumption $R_t^{ur}$ and for exploration $R_t^{ue}$ recommendations, respectively. 
\begin{align}
    \langle u, t, (U\times I)_{\leq t}, I_{\leq t}^u &\rangle \rightarrow R_t^{ur} \label{eqn:repeat}   \\
     \langle u, t, (U\times I)_{\leq t}, I_{\leq t}^{\bar{u}} &\rangle \rightarrow R_t^{ue} \label{eqn:explore}
\end{align}
Note that, how to present $R_t^{ur}$ and $R_t^{ue}$ to users, \eg as two separate rankings or as a merged ranking of items from both recommendations, is orthogonal to our discussion here.

The partitioning of candidate items  into $I^u$ and $I^{\bar{u}}$ may again seem trivial (the subscript $t$ is omitted for clarity), but it leads to (i) significant differences in the recommender's search space, since  $|I^u| \ll |I^{\bar{u}}|$ and $|I^{\bar{u}}| \approx |I|$, and (ii) consequently, notable impacts on evaluation. Because the recommendation space is much smaller and largely composed of earlier preferred items, making good recommendations among repeated items is far easier than selecting from a large pool of unfamiliar ones. For instance, in food delivery recommendation, the proposed model in~\cite{Jiayu24} achieved NDCG scores ranging from 0.59 to 0.64 for repeat consumption, while for exploration they ranged only from 0.09 to 0.17 across datasets from three cities. Similar performance gaps were observed for all evaluated baselines~\cite{Jiayu24}. The recommendation accuracy could be easily dominated by the repeated items if repeated consumption is common in the task setting, like food ordering and groceries shopping. Hence, the separation of evaluations of repeated consumption and exploration better reflects the model's accuracy. 

The repeated consumption and exploration is just one example of candidate item selection. The candidate items suitable for recommendation can be directly specified by users \eg by time, location, or any other item attributes. The key message here is that the set of eligible items to be recommended is determined \textit{before} running the recommender system.

\subsection{Task Formulation with Constraints}
We can now define a more general formulation by introducing a selection condition on the candidate items, denoted as $s(I_{\leq t})$. Specifically, $s(\cdot)$ represents selection criteria derived from a user's interaction history like repeated consumption, or be specified by the user through temporal, spatial, or other attribute-based constraints. By including $s(I_{\leq t})$ as part of the task input, the candidate set of items is determined prior to model ranking. In fact, candidate generation (also known as recall) constitutes the first stage of the multi-stage recommender system architecture widely adopted by today's largest online platforms, preceding the ranking and re-ranking stages~\cite{Reranking22}.

\begin{definition}[Recommendation Task]
\label{def:final}
Let $U$ be a set of users and $I$ be a set of items. A recommender system aims to produce a ranked list of items for a user $u$ at time $t$, based on user-item interactions $U \times I$ that is available at $t$, and the conditions specified on the candidate items. 
\begin{equation}\label{eqn:selection}
\left\langle u, t, (U\times I)_{\leq t}, s(I_{\leq t}) \right\rangle \rightarrow R_t^u    
\end{equation}
\end{definition}

The final items recommended to a user could be the results of running \textit{multiple recommendations made from multiple sets of candidate items} conditioned on different multiple $s(\cdot)$'s. In terms of model evaluation, it is more meaningful to independently evaluate each model specific to candidate items on one selection function $s(\cdot)$.

%====================
\section{From Recommendation to User Consumption}
\label{sec:interaction}
%====================

Users may interact with items for various reasons. An interaction could be the result of an intentional search rather than recommendation. For instance, the Yambda-5B dataset includes an \texttt{is\_organic} flag to indicate actions not driven by recommendations~\cite{Yambda5b25}. In this discussion, we generally assume that \textit{a user's interaction with an item reflects some degree of preference for it}.

Our focus is to examine the stages from a recommended item $i \in R_t^u$ to an interaction $(u, i, t_x)$, and to what extent a model can learn from such interactions. This process can be considered as another form a mapping:
\begin{equation}\label{eqn:interaction}
    \langle u, R_t^u\rangle \rightarrow (u, i, t_x) \text{\quad where \quad} i\in R_t^u
\end{equation}

%====================
\subsection{The Life Cycle of User-Item Interaction}
%====================
We divide user-item interactions into three stages: \textit{pre-interaction judgment}, \textit{interaction}, and \textit{post-interaction feedback}. These stages may not apply to all recommendation scenarios, but offer a useful framework for understanding the relationship between user interactions and preferences. A conceptually similar framework was proposed by~\citet{LeeUninteresting23}, who distinguish between \textit{pre-use preference} (a user's impression of an item before interacting with it) and \textit{post-use preference} (the impression formed after interaction). Our discussion  aims to highlight the different types of pre-interaction and interaction across various recommendation scenarios, which differs from both the technical solutions presented in~\cite{LeeUninteresting23} and the decision-making perspectives discussed in~\cite{handbookDecisionMaking}.

Consider a user booking a hotel in an unfamiliar city for a conference. When she first browses a list of options, she forms a \textit{pre-interaction judgment} based on visible cues such as images, branding, location, and price. After clicking on one hotel, she enters the deliberate evaluation stage --- reading descriptions, checking facilities, and reviewing feedback before booking. Months later, her stay in the hotel marks the \textit{interaction} stage. After checkout, she leaves a rating and review, marking the \textit{post-interaction feedback} stage.

Both quick judgments and deliberate evaluations in the pre-interaction stage come at a cost, the user's time and effort in gathering relevant information. The outcomes of these efforts typically indicate user preference, especially in similar future situations, \eg planning another trip. 
Post-interaction feedback, however, may not always accurately reflect user preference. While a review can highlight the primary reasons for choosing a hotel, representing genuine user preference, it may also describe aspects such as expectation gaps and booking effort, which are not necessarily preference indicators. A rating, influenced by these factors, may not always be a reliable measure of \textit{preference} compared to the fact that the user chose to stay at the hotel.  In many practical scenarios, post-interaction feedback is often difficult to collect, as it requires users to put in additional effort. Hence, in our following discussion, we focus on the other two stages.

%===========================
\subsection{Complexity of Pre-Interaction Judgment}
%===========================

The complexity of pre-interaction judgment from the user's perspective may arise from several dimensions.

\subsubsection{Informed vs Uninformed Decision} One major factor from the user's perspective is whether they possess the knowledge to accurately judge an item before interacting with it. For familiar items like books, movies, or other products the user has prior experience with, they can make an informed decision based on available attributes or other relevant information. Take movies as an example, users may decide whether to watch a film based on the director, cast, genre, a brief synopsis, or simply the poster. However, if a user has never used a robot vacuum before, many of the terms in the product description may be unfamiliar to them. Even after reading the machine specifications and user reviews, she may struggle to discern the pros and cons of a specific model, with references to her own home layout and floor conditions.

Uninformed decisions may not necessarily indicate user preferences. It is also challenging for a recommendation platform to determine whether a user's decision was based on prior knowledge or made without a full understanding of the item. This distinction is crucial, as recommendations based on uninformed interactions might not truly reflect user interests, potentially leading to less effective personalization. For example, after using a robot vacuum for some time, the user may realize that she should purchase another model with features better suited to her floor conditions.

\subsubsection{Items in One vs. Multiple Types} Many studies in recommendation are conducted on datasets containing only one type of item, such as books, movies, news, or music. In these cases, there are common characteristics, such as genre, director, or artist, that users can rely on for pre-interaction judgment before actually interacting with a recommended item. However, in the context of online shopping, where products span thousands of categories, users apply different criteria and expectations when making judgments for different types of items. 

In such a diverse setting, user preferences learned from interactions across multiple item types may be shaped more by general associative patterns than by genuine preferences for specific products. However, distinguishing between personal preferences and co-purchasing patterns remains challenging, as these so-called preferences may be expressed for broader item categories rather than specific items. Nevertheless, we acknowledge that user preference is a complex concept~\cite{DecisionMakingEC22}.

%===========================
\subsection{Recognition of User-Item Interaction}
%===========================
User-item interactions are often recorded in the form of $(u, i, t_x)$ in a RecSys dataset. However, interactions may appear in different forms depending on the recommendation context.

\subsubsection{Interaction Process Can Be Complicated.} Taking online shopping as an example, the process does not end when a user clicks on a recommended item. After deciding on a product, the user might add the item to their cart, proceed to payment, receive the delivery, and ultimately complete the purchase. From the shopping platform's perspective, this sequence of events marks a successful interaction. However, complications can arise if the user later decides to return the product due to reasons such as quality issues, unmet expectations, or simply a change of mind.

This raises an important question: should an unsuccessful purchase (\ie one that results in a return) still be considered a valid user-item interaction for learning user preferences? On the one hand, the initial decision to purchase the item indicates some level of interest or preference. On the other hand, the return suggests dissatisfaction or misalignment with the user's expectations. If returns are not accounted for, the model might incorrectly reinforce recommendations for similar items, leading to suboptimal suggestions. Therefore, when incorporating interaction data into preference modeling, it is crucial to differentiate between successful and unsuccessful purchases and consider additional contextual signals, such as return rates, to better understand user preferences.

\subsubsection{Interaction without Pre-Interaction Judgment.} Some user-item interactions happen without a pre-interaction judgment phase. This is especially common in scenarios like music streaming and short-video viewing, where users often do not actively select each item. Instead, they are presented with an initial set of options, and after selecting the first item, subsequent content is automatically fed to users by the recommendation engine, \eg a playlist or streaming.

In such cases, user engagement signals, such as skipping, fast-forwarding, or continuing to watch/listen, play a crucial role in modeling preferences. Unlike traditional recommendation settings where users consciously evaluate and select items before interaction, here, user preferences are inferred more dynamically based on real-time behaviors, or the interaction itself. Recommendation models must distinguish between passive exposure and active preference while adapting to continuously evolving user interests. In this case, the common form of user-item interaction $(u, i, t_x)$ becomes less accurate compared to other settings. Such recommendation in streaming form also post questions in item attribute modeling \eg evaluating whether the cover image of a short video influence user viewing as the user may not even has the chance of viewing the cover image for each video in the streaming. 

\subsubsection{Unobservable Interaction.} There are also recommendation scenarios where user-item interactions are not directly observable and can only be inferred from external sources.

One example is job recommendation, where matching is based on a user's skills and knowledge against job requirements. Unlike traditional recommender systems where user engagement signals (such as clicks, purchases, or views) are readily available, the job application process involves significant effort on both ends, applicants must prepare resumes and cover letters, while companies conduct interviews before making offers. As a result, direct interactions, such as applying for a job or receiving an offer, may not always be captured by a job recommendation platform. Instead, implicit signals, such as a user frequently viewing job postings in a particular field or updating their profile, might be used to infer their interests and preferences.

This lack of direct interaction data introduces challenges in preference modeling, as user engagement may not always reflect strong interest, and external factors \eg hiring decisions, can influence the outcome. Thus, in such cases, recommender systems must rely on richer contextual information and alternative feedback mechanisms to refine their predictions.

%====================
\subsection{Interdependency across Recommendations}
%====================
Recommendations can occur either independently, as in hotel booking, or within a session-based context, such as music streaming and short-video viewing. In the latter case, subsequent recommendations may be influenced by the previous selections or even the initial choice made at the start of the session.
This is a key characteristic of session-based recommendation, a specialized type of recommendation task~\cite{YaoSurvey25}. A detailed discussion on the recommendation flow is  made in~\cite{SunBeyond24}.

Each recommendation scenario, whether based on user preferences, session context, or product types, requires careful formulation to ensure that the user experience is optimized and that the system accurately reflects user intent.

%====================
\section{Discussion and Perspectives}
\label{sec:discussion}
%====================
Next, we discuss the ultimate goal of recommendation, examine the relationships among task formulation, solution, and evaluation, explore the intersection of task specificity and model generalizability, and provide actionable guidance for RecSys academic research.

%====================
\subsection{Recommendation vs User Cost}
%====================

The ultimate goal of a recommender system is to reduce user effort in finding products or services of interest, enhance their enjoyment of recommendations, and build trust in the system. However, users still incur different costs at various stages of the interaction process. In the \textit{pre-interaction judgment} stage, users  evaluate recommendations based on available information, such as descriptions, images, reviews, or other metadata, which requires cognitive effort and time. During the \textit{interaction} stage, users experience the actual product or service, which may involve monetary costs (\eg purchasing a product), time investment (\eg watching a recommended movie), or engagement effort (\eg exploring an unfamiliar interface). If the recommendation is poor, users may feel frustrated, leading to dissatisfaction and disengagement~\cite{Hesitation24}. Finally, in the \textit{post-interaction} stage, users may be asked to provide feedback, such as ratings or reviews. This step requires additional effort, and many users may choose not to participate. 

Understanding and minimizing these costs at each of the three stages is essential for improving user experience and optimizing the effectiveness, and even the trustworthiness, of recommender systems. However, as illustrated in the examples above, such costs manifest differently across recommendation scenarios. The RecSys task definition in Equation~\ref{eqn:selection} primarily specifies the \textit{inputs} a recommender model should consider and the desired \textit{output}. Yet, the various costs incurred by users at different stages from recommendation to interaction (Equation~\ref{eqn:interaction}) cannot be explicitly represented in a formal formulation. Nevertheless, understanding these costs can inform the design of more refined loss functions for learning recommendation models. For example, in short-video recommendation, videos that a user watches for a duration shorter than their average viewing percentage can be treated as tolerance samples. In~\cite{Hesitation24}, such samples are assigned different losses compared to fully watched videos, leading to improvements in user retention verified through A/B testing. In this context, watching an uninteresting video constitutes a cost to the user.

%====================
\subsection{Task, Solution, and Evaluation}
\label{ssec:discussionTask}
%====================

A task formulation is often a formal abstraction of real-world applications. While such abstractions may omit details specific to certain recommendation platforms, the solutions developed should remain confined to the defined input and output of the task formulation. Consequently, since offline evaluation often serves as a proxy for selecting the most promising solutions for online evaluation, it should be designed to assess a solution's performance with respect to the task formulation itself. 

In RecSys research, evaluation is sometimes tailored to fit the proposed solutions rather than being aligned with the task formulation. One example is the evaluation of Bandits and reinforcement learning-based RecSys solutions. As reported in~\cite{FreshSun23}, when examining dataset partitioning methods, it was observed that only a small number of papers followed the timeline and simulated user interactions over time, which is a good practice. However, the evaluation was not motivated by the task itself, but because reinforcement learning solutions require reward signals based on  user actions, leading to an evaluation setup that caters to the proposed solution.

Some existing RecSys tasks are not well formulated. Sequential recommendation and intent-aware recommendation are two examples; they do not fundamentally differ from a typical recommendation problem. In the survey paper, \citet{IntentSurvey24} defines: intent-aware recommender ``is a recommender system that is designed to capture the users' underlying current motivations and goals in order to support them.'' If we view the problem definition by its inputs and outputs, there is no significant difference from the definition in Equation~\ref{eqn:selection}. Probably due to a less distinctive problem formulation, there are questions on the progress made~\cite{IntentawareWorrying25}. In the survey paper~\cite{SurveySequence24}, the task of sequential recommendation is defined as follows: ``In sequential recommendation, we often have one or more sequences of interacted items w.r.t. each user, as well as some auxiliary information to help learn user preferences. Our goal is then to generate a ranked list of items accurately for each user''. This is basically a different view of the very same $U \times I$. Naturally all items interacted with by a user form a sequence as illustrated in Figure~\ref{fig:taskTime}. Solutions that pay attention on such sequences may be able to achieve a better recommendation accuracy, particularly in the RecSys settings with items of a single type \eg music, movie, book, and news. However, the task to be addressed remains a generic recommendation.

%====================
\subsection{Task Specificity vs Model Generalizability}
\label{ssec:specifity}
%====================

Recommender systems represent a fascinating intersection between theoretical research and practical deployment. On the one hand, they are directly linked to real-world applications, where performance improvements can lead to tangible business benefits and enhanced user experiences. On the other hand, academic research often strives to develop generic models that generalize well across diverse datasets and application scenarios. The tension between these two objectives creates an interesting challenge: we need models that perform well across multiple datasets representing different recommendation settings, yet optimizing a model for a specific application often requires customization in terms of input features, objective functions, and even model architecture.

This trade-off between task specificity and model generalizability is a fundamental issue in recommender system research. Highly specialized models, fine-tuned for a particular domain, can achieve state-of-the-art performance in their respective tasks but may struggle when applied to other recommendation settings. Conversely, more general models that are designed to work across various domains often sacrifice performance in any given task, as they cannot fully exploit the domain-specific characteristics that drive user preferences. This issue is further exacerbated by the diverse nature of recommendation tasks, as discussed earlier.

%=========================
\subsection{From Theory to Practice}
\label{ssec:actionable}
%=========================

Moving forward, a practical step toward addressing the challenges discussed earlier is to establish a clear categorization of recommendation tasks. In our recent survey~\cite{zou2025survey}, which includes only research papers reporting online A/B testing results from production environments, we classify \textit{real-world recommendation} tasks into two main categories.   \textit{Transaction-oriented recommender systems} generate item recommendations with the primary goal of prompting transactional user actions, optimizing for metrics such as conversion rate, revenue, or purchase likelihood. E-commerce platforms are typical examples.
\textit{Content-oriented recommender systems}, on the other hand, generate recommendations to facilitate user consumption and engagement, optimizing for metrics such as dwell time, clicks, or user satisfaction. Common examples include news, video, and music recommenders. Although these two main categories do not account for all types of recommendation systems, they capture the majority of widely studied recommendation scenarios.

The objectives of recommender systems across these two broad categories differ substantially and can vary further within each category. Consequently, more specific recommendation tasks can be refined along key dimensions discussed earlier. For instance, recency is a key factor in news recommendation but may be less relevant for other types of content-oriented recommendation. RecSys research focused on algorithmic advancement should clearly specify the recommendation task it addresses, use datasets and evaluation settings that reflect realistic conditions, and compare against baselines designed for similar contexts. 

Such categorization ensures fair and meaningful evaluation across different recommendation tasks. Importantly, this emphasis on task specificity also requires a shared understanding within the community, especially among reviewers, that researchers should not be expected to overgeneralize their solutions to unrelated datasets or problem settings. For example, a submission should not be penalized for not reporting results on widely used datasets like MovieLens, when it addresses fundamentally different recommendation scenarios. The effectiveness of a recommender system depends much on how well it aligns with the specific characteristics of the target task.

We highlight two recent examples that suggest researchers and reviewers are increasingly moving in this direction.  \citet{Playlist25} propose a playlist-generation recommender focused on a specific task: generating playlists from titles. Their method is evaluated on the Million Playlist Dataset and compared against task-relevant baselines. The authors further analyze user effort in playlist creation and assess the usefulness of playlist titles, demonstrating strong task alignment rather than pursuing cross-domain generalization. Similarly, based on their experience building a small-scale production news recommender system, \citet{higley2025news} observe that “published models are surprisingly difficult to apply to the kinds of data found in real-world datasets and practical recommendation problems.” They emphasize that “building recommender systems that serve real users requires deep and specific engagement not just with a domain in general, but with the particular characteristics of specific datasets, applications, and user communities,” underscoring the task-specific nature of RecSys research. 
Task-specific RecSys research has become increasingly feasible thanks to dataset availability. For example, Yambda-5B includes both implicit (\eg listening) and explicit (\eg likes and dislikes) feedback, as well as organic user actions specific to the music streaming recommendation task~\cite{Yambda5b25}. The authors also introduce a global temporal split to better reflect real-world settings and prevent
data leakage.

%====================
\section{Conclusion}
\label{sec:conclude}
%====================
While foundational works in recommender systems are well established, there remains a lack of consensus within the community regarding baseline models and datasets. The reason behind is likely stemming from an insufficient understanding of RecSys task formulation. In this paper, we provide an in-depth analysis of recommendation tasks, emphasizing the need for clear and well-defined problem definitions to enable effective evaluation and the development of more practical solutions. We highlight the importance of understanding input-output relationships in recommendation models, accounting for factors such as temporal dynamics and candidate item selection. We further examine the complexities of user-item interactions, including user-incurred costs during decision-making and the challenges introduced by multi-step interactions in real-world settings.  Ultimately, we argue that the central objective of RecSys is to minimize user costs across the entire recommendation process. The nature of these costs provides a basis for distinguishing among different recommendation tasks.

This paper aims to clarify the relationship between task formulation, solution design, and evaluation, particularly offline evaluation. By promoting a more structured and comprehensive understanding of these key elements, we hope to deepen the community's appreciation of RecSys task complexity and support both new and experienced researchers in advancing the field across diverse real-world domains. We also urge researchers and reviewers to recognize the importance of task specificity and to value innovations tailored to distinct recommendation scenarios.

%% the bibliography file.
\bibliographystyle{ACM-Reference-Format}
\bibliography{RecSysBib}

%%% -*-BibTeX-*-
%%% Do NOT edit. File created by BibTeX with style
%%% ACM-Reference-Format-Journals [18-Jan-2012].

\begin{thebibliography}{35}

%%% ====================================================================
%%% NOTE TO THE USER: you can override these defaults by providing
%%% customized versions of any of these macros before the \bibliography
%%% command.  Each of them MUST provide its own final punctuation,
%%% except for \shownote{}, \showDOI{}, and \showURL{}.  The latter two
%%% do not use final punctuation, in order to avoid confusing it with
%%% the Web address.
%%%
%%% To suppress output of a particular field, define its macro to expand
%%% to an empty string, or better, \unskip, like this:
%%%
%%% \newcommand{\showDOI}[1]{\unskip}   % LaTeX syntax
%%%
%%% \def \showDOI #1{\unskip}           % plain TeX syntax
%%%
%%% ====================================================================

\ifx \showCODEN    \undefined \def \showCODEN     #1{\unskip}     \fi
\ifx \showDOI      \undefined \def \showDOI       #1{#1}\fi
\ifx \showISBNx    \undefined \def \showISBNx     #1{\unskip}     \fi
\ifx \showISBNxiii \undefined \def \showISBNxiii  #1{\unskip}     \fi
\ifx \showISSN     \undefined \def \showISSN      #1{\unskip}     \fi
\ifx \showLCCN     \undefined \def \showLCCN      #1{\unskip}     \fi
\ifx \shownote     \undefined \def \shownote      #1{#1}          \fi
\ifx \showarticletitle \undefined \def \showarticletitle #1{#1}   \fi
\ifx \showURL      \undefined \def \showURL       {\relax}        \fi
% The following commands are used for tagged output and should be
% invisible to TeX
\providecommand\bibfield[2]{#2}
\providecommand\bibinfo[2]{#2}
\providecommand\natexlab[1]{#1}
\providecommand\showeprint[2][]{arXiv:#2}

\bibitem[Adomavicius and Tuzhilin(2005)]%
        {TkdeSurvey05}
\bibfield{author}{\bibinfo{person}{Gediminas Adomavicius} {and} \bibinfo{person}{Alexander Tuzhilin}.} \bibinfo{year}{2005}\natexlab{}.
\newblock \showarticletitle{Toward the Next Generation of Recommender Systems: {A} Survey of the State-of-the-Art and Possible Extensions}.
\newblock \bibinfo{journal}{\emph{{IEEE} {TKDE}}} \bibinfo{volume}{17}, \bibinfo{number}{6} (\bibinfo{year}{2005}), \bibinfo{pages}{734--749}.
\newblock
\urldef\tempurl%
\url{https://doi.org/10.1109/TKDE.2005.99}
\showDOI{\tempurl}


\bibitem[Anderson et~al\mbox{.}(2014)]%
        {anderson2014dynamics}
\bibfield{author}{\bibinfo{person}{Ashton Anderson}, \bibinfo{person}{Ravi Kumar}, \bibinfo{person}{Andrew Tomkins}, {and} \bibinfo{person}{Sergei Vassilvitskii}.} \bibinfo{year}{2014}\natexlab{}.
\newblock \showarticletitle{The dynamics of repeat consumption}. In \bibinfo{booktitle}{\emph{WWW}}. \bibinfo{pages}{419--430}.
\newblock


\bibitem[Ariannezhad et~al\mbox{.}(2022)]%
        {ariannezhad2022recanet}
\bibfield{author}{\bibinfo{person}{Mozhdeh Ariannezhad}, \bibinfo{person}{Sami Jullien}, \bibinfo{person}{Ming Li}, \bibinfo{person}{Min Fang}, \bibinfo{person}{Sebastian Schelter}, {and} \bibinfo{person}{Maarten de Rijke}.} \bibinfo{year}{2022}\natexlab{}.
\newblock \showarticletitle{ReCANet: A repeat consumption-aware neural network for next basket recommendation in grocery shopping}. In \bibinfo{booktitle}{\emph{ACM SIGIR}}. \bibinfo{pages}{1240--1250}.
\newblock


\bibitem[Beel et~al\mbox{.}(2024)]%
        {DatasetRecsys24}
\bibfield{author}{\bibinfo{person}{Joeran Beel}, \bibinfo{person}{Lukas Wegmeth}, \bibinfo{person}{Lien Michiels}, {and} \bibinfo{person}{Steffen Schulz}.} \bibinfo{year}{2024}\natexlab{}.
\newblock \showarticletitle{Informed Dataset Selection with ‘Algorithm Performance Spaces’}. In \bibinfo{booktitle}{\emph{ACM RecSys}}. \bibinfo{publisher}{{ACM}}, \bibinfo{pages}{1085–1090}.
\newblock
\showISBNx{9798400705052}
\urldef\tempurl%
\url{https://doi.org/10.1145/3640457.3691704}
\showDOI{\tempurl}


\bibitem[Charolois-Pasqua et~al\mbox{.}(2025)]%
        {Playlist25}
\bibfield{author}{\bibinfo{person}{Enzo Charolois-Pasqua}, \bibinfo{person}{El\'{e}a Vellard}, \bibinfo{person}{Youssra Rebboud}, \bibinfo{person}{Pasquale Lisena}, {and} \bibinfo{person}{Rapha\"{e}l Troncy}.} \bibinfo{year}{2025}\natexlab{}.
\newblock \showarticletitle{A Language Model-Based Playlist Generation Recommender System}. In \bibinfo{booktitle}{\emph{{ACM} {RecSys}}} \emph{(\bibinfo{series}{RecSys '25})}. \bibinfo{publisher}{ACM}, \bibinfo{address}{New York, NY, USA}, \bibinfo{pages}{1–11}.
\newblock
\showISBNx{9798400713644}
\urldef\tempurl%
\url{https://doi.org/10.1145/3705328.3748053}
\showDOI{\tempurl}


\bibitem[Herlocker et~al\mbox{.}(2004)]%
        {HerlockerKTR04}
\bibfield{author}{\bibinfo{person}{Jonathan~L. Herlocker}, \bibinfo{person}{Joseph~A. Konstan}, \bibinfo{person}{Loren~G. Terveen}, {and} \bibinfo{person}{John Riedl}.} \bibinfo{year}{2004}\natexlab{}.
\newblock \showarticletitle{Evaluating collaborative filtering recommender systems}.
\newblock \bibinfo{journal}{\emph{{ACM} {TOIS}}} \bibinfo{volume}{22}, \bibinfo{number}{1} (\bibinfo{year}{2004}), \bibinfo{pages}{5--53}.
\newblock
\urldef\tempurl%
\url{https://doi.org/10.1145/963770.963772}
\showDOI{\tempurl}


\bibitem[Higley et~al\mbox{.}(2025)]%
        {higley2025news}
\bibfield{author}{\bibinfo{person}{Karl Higley}, \bibinfo{person}{Robin Burke}, \bibinfo{person}{Michael~D. Ekstrand}, {and} \bibinfo{person}{Bart~P. Knijnenburg}.} \bibinfo{year}{2025}\natexlab{}.
\newblock \showarticletitle{What News Recommendation Research Did (But Mostly Didn't) Teach Us About Building A News Recommender}. In \bibinfo{booktitle}{\emph{Proc. of BEYOND 2025 Workshop co-located ACM RecSys}}.
\newblock


\bibitem[Ivanova et~al\mbox{.}(2023)]%
        {RecBaselines23}
\bibfield{author}{\bibinfo{person}{Veronika Ivanova}, \bibinfo{person}{Oleg Lashinin}, \bibinfo{person}{Marina Ananyeva}, {and} \bibinfo{person}{Sergey Kolesnikov}.} \bibinfo{year}{2023}\natexlab{}.
\newblock \showarticletitle{{RecBaselines2023}: a new dataset for choosing baselines for recommender models}. In \bibinfo{booktitle}{\emph{{Workshop} on {Bibliometric}-enhanced {Information} {Retrieval}}} \emph{(\bibinfo{series}{{CEUR} {Workshop} {Proceedings}}, Vol.~\bibinfo{volume}{3617})}. \bibinfo{publisher}{CEUR}, \bibinfo{pages}{52--65}.
\newblock
\newblock
\shownote{ISSN: 1613-0073}.


\bibitem[Jameson et~al\mbox{.}(2022)]%
        {handbookDecisionMaking}
\bibfield{author}{\bibinfo{person}{Anthony Jameson}, \bibinfo{person}{Martijn~C. Willemsen}, {and} \bibinfo{person}{Alexander Felfernig}.} \bibinfo{year}{2022}\natexlab{}.
\newblock \bibinfo{booktitle}{\emph{Individual and Group Decision Making and Recommender Systems}}.
\newblock \bibinfo{publisher}{Springer US}, \bibinfo{address}{New York, NY}, \bibinfo{pages}{789--832}.
\newblock
\showISBNx{978-1-0716-2197-4}
\urldef\tempurl%
\url{https://doi.org/10.1007/978-1-0716-2197-4_21}
\showDOI{\tempurl}


\bibitem[Jannach and Zanker(2024)]%
        {IntentSurvey24}
\bibfield{author}{\bibinfo{person}{Dietmar Jannach} {and} \bibinfo{person}{Markus Zanker}.} \bibinfo{year}{2024}\natexlab{}.
\newblock \showarticletitle{A Survey on Intent-aware Recommender Systems}.
\newblock \bibinfo{journal}{\emph{ACM TORS}} \bibinfo{volume}{3}, \bibinfo{number}{2}, Article \bibinfo{articleno}{23} (\bibinfo{date}{Dec.} \bibinfo{year}{2024}), \bibinfo{numpages}{32}~pages.
\newblock
\urldef\tempurl%
\url{https://doi.org/10.1145/3700890}
\showDOI{\tempurl}


\bibitem[Ji et~al\mbox{.}(2020)]%
        {PopularitySigir20}
\bibfield{author}{\bibinfo{person}{Yitong Ji}, \bibinfo{person}{Aixin Sun}, \bibinfo{person}{Jie Zhang}, {and} \bibinfo{person}{Chenliang Li}.} \bibinfo{year}{2020}\natexlab{}.
\newblock \showarticletitle{A Re-visit of the Popularity Baseline in Recommender Systems}. In \bibinfo{booktitle}{\emph{{ACM} {SIGIR}}}. \bibinfo{pages}{1749--1752}.
\newblock
\urldef\tempurl%
\url{https://doi.org/10.1145/3397271.3401233}
\showDOI{\tempurl}


\bibitem[Ji et~al\mbox{.}(2023)]%
        {DataLeakage23}
\bibfield{author}{\bibinfo{person}{Yitong Ji}, \bibinfo{person}{Aixin Sun}, \bibinfo{person}{Jie Zhang}, {and} \bibinfo{person}{Chenliang Li}.} \bibinfo{year}{2023}\natexlab{}.
\newblock \showarticletitle{A Critical Study on Data Leakage in Recommender System Offline Evaluation}.
\newblock \bibinfo{journal}{\emph{{ACM} {TOIS}}} \bibinfo{volume}{41}, \bibinfo{number}{3} (\bibinfo{year}{2023}), \bibinfo{pages}{75:1--75:27}.
\newblock
\urldef\tempurl%
\url{https://doi.org/10.1145/3569930}
\showDOI{\tempurl}


\bibitem[Katz et~al\mbox{.}(2022)]%
        {katz2022learning}
\bibfield{author}{\bibinfo{person}{Ori Katz}, \bibinfo{person}{Oren Barkan}, \bibinfo{person}{Noam Koenigstein}, {and} \bibinfo{person}{Nir Zabari}.} \bibinfo{year}{2022}\natexlab{}.
\newblock \showarticletitle{Learning to Ride a Buy-Cycle: A Hyper-Convolutional Model for Next Basket Repurchase Recommendation}. In \bibinfo{booktitle}{\emph{ACM RecSys}}. \bibinfo{pages}{316--326}.
\newblock


\bibitem[Kleinberg et~al\mbox{.}(2022)]%
        {DecisionMakingEC22}
\bibfield{author}{\bibinfo{person}{Jon~M. Kleinberg}, \bibinfo{person}{Sendhil Mullainathan}, {and} \bibinfo{person}{Manish Raghavan}.} \bibinfo{year}{2022}\natexlab{}.
\newblock \showarticletitle{The Challenge of Understanding What Users Want: Inconsistent Preferences and Engagement Optimization}. In \bibinfo{booktitle}{\emph{{ACM} Conference on Economics and Computation (EC)}}.
\newblock
\urldef\tempurl%
\url{https://doi.org/10.1145/3490486.3538365}
\showDOI{\tempurl}


\bibitem[Koren et~al\mbox{.}(2009)]%
        {KorenBV09}
\bibfield{author}{\bibinfo{person}{Yehuda Koren}, \bibinfo{person}{Robert~M. Bell}, {and} \bibinfo{person}{Chris Volinsky}.} \bibinfo{year}{2009}\natexlab{}.
\newblock \showarticletitle{Matrix Factorization Techniques for Recommender Systems}.
\newblock \bibinfo{journal}{\emph{Computer}} \bibinfo{volume}{42}, \bibinfo{number}{8} (\bibinfo{year}{2009}), \bibinfo{pages}{30--37}.
\newblock
\urldef\tempurl%
\url{https://doi.org/10.1109/MC.2009.263}
\showDOI{\tempurl}


\bibitem[Le et~al\mbox{.}(2025)]%
        {Le25Dont}
\bibfield{author}{\bibinfo{person}{Huy~Hoang Le}, \bibinfo{person}{Yang Liu}, \bibinfo{person}{Alan Medlar}, {and} \bibinfo{person}{Dorota Glowacka}.} \bibinfo{year}{2025}\natexlab{}.
\newblock \showarticletitle{Don't Get Ahead of Yourself: A Critical Study on Data Leakage in Offline Evaluation of Sequential Recommenders}. In \bibinfo{booktitle}{\emph{{ACM} {RecSys}}} \emph{(\bibinfo{series}{RecSys '25})}. \bibinfo{publisher}{ACM}, \bibinfo{pages}{1164–1168}.
\newblock
\showISBNx{9798400713644}
\urldef\tempurl%
\url{https://doi.org/10.1145/3705328.3759329}
\showDOI{\tempurl}


\bibitem[Lee and Kim(2023)]%
        {LeeUninteresting23}
\bibfield{author}{\bibinfo{person}{Yeon-Chang Lee} {and} \bibinfo{person}{Sang-Wook Kim}.} \bibinfo{year}{2023}\natexlab{}.
\newblock \showarticletitle{Uninteresting Items: Concept and Its Application to Effective Collaborative Filtering in Recommender Systems}.
\newblock \bibinfo{journal}{\emph{SIGWEB Newsl.}} \bibinfo{volume}{2023}, \bibinfo{number}{Autumn}, Article \bibinfo{articleno}{4} (\bibinfo{date}{Dec.} \bibinfo{year}{2023}), \bibinfo{numpages}{13}~pages.
\newblock
\showISSN{1931-1745}
\urldef\tempurl%
\url{https://doi.org/10.1145/3631358.3631362}
\showDOI{\tempurl}


\bibitem[Li et~al\mbox{.}(2024)]%
        {Jiayu24}
\bibfield{author}{\bibinfo{person}{Jiayu Li}, \bibinfo{person}{Aixin Sun}, \bibinfo{person}{Weizhi Ma}, \bibinfo{person}{Peijie Sun}, {and} \bibinfo{person}{Min Zhang}.} \bibinfo{year}{2024}\natexlab{}.
\newblock \showarticletitle{Right Tool, Right Job: Recommendation for Repeat and Exploration Consumption in Food Delivery}. In \bibinfo{booktitle}{\emph{{ACM} {RecSys}}}. \bibinfo{pages}{643--653}.
\newblock
\urldef\tempurl%
\url{https://doi.org/10.1145/3640457.3688119}
\showDOI{\tempurl}


\bibitem[Li et~al\mbox{.}(2025)]%
        {YaoSurvey25}
\bibfield{author}{\bibinfo{person}{Zihao Li}, \bibinfo{person}{Chao Yang}, \bibinfo{person}{Yakun Chen}, \bibinfo{person}{Xianzhi Wang}, \bibinfo{person}{Hongxu Chen}, \bibinfo{person}{Guandong Xu}, \bibinfo{person}{Lina Yao}, {and} \bibinfo{person}{Michael Sheng}.} \bibinfo{year}{2025}\natexlab{}.
\newblock \showarticletitle{Graph and Sequential Neural Networks in Session-based Recommendation: {A} Survey}.
\newblock \bibinfo{journal}{\emph{{ACM} Comput. Surv.}} \bibinfo{volume}{57}, \bibinfo{number}{2} (\bibinfo{year}{2025}), \bibinfo{pages}{40:1--40:37}.
\newblock
\urldef\tempurl%
\url{https://doi.org/10.1145/3696413}
\showDOI{\tempurl}


\bibitem[Liu et~al\mbox{.}(2022)]%
        {Reranking22}
\bibfield{author}{\bibinfo{person}{Weiwen Liu}, \bibinfo{person}{Yunjia Xi}, \bibinfo{person}{Jiarui Qin}, \bibinfo{person}{Fei Sun}, \bibinfo{person}{Bo Chen}, \bibinfo{person}{Weinan Zhang}, \bibinfo{person}{Rui Zhang}, {and} \bibinfo{person}{Ruiming Tang}.} \bibinfo{year}{2022}\natexlab{}.
\newblock \showarticletitle{Neural Re-ranking in Multi-stage Recommender Systems: {A} Review}. In \bibinfo{booktitle}{\emph{{IJCAI}}}. \bibinfo{publisher}{ijcai.org}, \bibinfo{pages}{5512--5520}.
\newblock
\urldef\tempurl%
\url{https://doi.org/10.24963/IJCAI.2022/771}
\showDOI{\tempurl}


\bibitem[McElfresh et~al\mbox{.}(2022)]%
        {McElfreshKV0W22}
\bibfield{author}{\bibinfo{person}{Duncan~C. McElfresh}, \bibinfo{person}{Sujay Khandagale}, \bibinfo{person}{Jonathan Valverde}, \bibinfo{person}{John Dickerson}, {and} \bibinfo{person}{Colin White}.} \bibinfo{year}{2022}\natexlab{}.
\newblock \showarticletitle{On the Generalizability and Predictability of Recommender Systems}. In \bibinfo{booktitle}{\emph{NeurIPS}}.
\newblock


\bibitem[Pan et~al\mbox{.}(2026)]%
        {SurveySequence24}
\bibfield{author}{\bibinfo{person}{Li-Wei Pan}, \bibinfo{person}{Wei-Ke Pan}, \bibinfo{person}{Mei-Yan Wei}, \bibinfo{person}{Hong-Zhi Yin}, {and} \bibinfo{person}{Zhong Ming}.} \bibinfo{year}{2026}\natexlab{}.
\newblock \showarticletitle{A survey on sequential recommendation}.
\newblock \bibinfo{journal}{\emph{Frontiers of Computer Science}}  \bibinfo{volume}{20} (\bibinfo{year}{2026}), \bibinfo{pages}{2003606--}.
\newblock
\showISSN{2095-2228}
\urldef\tempurl%
\url{https://doi.org/10.1007/s11704-025-41329-w}
\showDOI{\tempurl}


\bibitem[Ploshkin et~al\mbox{.}(2025)]%
        {Yambda5b25}
\bibfield{author}{\bibinfo{person}{Alexander Ploshkin}, \bibinfo{person}{Vladislav Tytskiy}, \bibinfo{person}{Alexey Pismenny}, \bibinfo{person}{Vladimir Baikalov}, \bibinfo{person}{Evgeny Taychinov}, \bibinfo{person}{Artem Permiakov}, \bibinfo{person}{Daniil Burlakov}, {and} \bibinfo{person}{Eugene Krofto}.} \bibinfo{year}{2025}\natexlab{}.
\newblock \showarticletitle{Yambda-5B — A Large-Scale Multi-Modal Dataset for Ranking and Retrieval}. In \bibinfo{booktitle}{\emph{{ACM} {RecSys}}} \emph{(\bibinfo{series}{RecSys '25})}. \bibinfo{publisher}{ACM}, \bibinfo{pages}{894–901}.
\newblock
\showISBNx{9798400713644}
\urldef\tempurl%
\url{https://doi.org/10.1145/3705328.3748163}
\showDOI{\tempurl}


\bibitem[Quan et~al\mbox{.}(2023)]%
        {quan2023enhancing}
\bibfield{author}{\bibinfo{person}{Shigang Quan}, \bibinfo{person}{Shui Liu}, \bibinfo{person}{Zhenzhe Zheng}, {and} \bibinfo{person}{Fan Wu}.} \bibinfo{year}{2023}\natexlab{}.
\newblock \showarticletitle{Enhancing Repeat-Aware Recommendation from a Temporal-Sequential Perspective}. In \bibinfo{booktitle}{\emph{ACM CIKM}}. \bibinfo{pages}{2095--2105}.
\newblock


\bibitem[Reiter-Haas et~al\mbox{.}(2021)]%
        {reiter2021predicting}
\bibfield{author}{\bibinfo{person}{Markus Reiter-Haas}, \bibinfo{person}{Emilia Parada-Cabaleiro}, \bibinfo{person}{Markus Schedl}, \bibinfo{person}{Elham Motamedi}, \bibinfo{person}{Marko Tkalcic}, {and} \bibinfo{person}{Elisabeth Lex}.} \bibinfo{year}{2021}\natexlab{}.
\newblock \showarticletitle{Predicting music relistening behavior using the ACT-R framework}. In \bibinfo{booktitle}{\emph{ACM RecSys}}. \bibinfo{pages}{702--707}.
\newblock


\bibitem[Rendle et~al\mbox{.}(2009)]%
        {ImplicitFeeback09}
\bibfield{author}{\bibinfo{person}{Steffen Rendle}, \bibinfo{person}{Christoph Freudenthaler}, \bibinfo{person}{Zeno Gantner}, {and} \bibinfo{person}{Lars Schmidt-Thieme}.} \bibinfo{year}{2009}\natexlab{}.
\newblock \showarticletitle{BPR: Bayesian personalized ranking from implicit feedback}. In \bibinfo{booktitle}{\emph{Conference on Uncertainty in Artificial Intelligence}}. \bibinfo{publisher}{AUAI Press}, \bibinfo{pages}{452–461}.
\newblock
\showISBNx{9780974903958}


\bibitem[Ricci et~al\mbox{.}(2011)]%
        {RicciRS11Chp}
\bibfield{author}{\bibinfo{person}{Francesco Ricci}, \bibinfo{person}{Lior Rokach}, {and} \bibinfo{person}{Bracha Shapira}.} \bibinfo{year}{2011}\natexlab{}.
\newblock \showarticletitle{Introduction to Recommender Systems Handbook}.
\newblock In \bibinfo{booktitle}{\emph{Recommender Systems Handbook}}, \bibfield{editor}{\bibinfo{person}{Francesco Ricci}, \bibinfo{person}{Lior Rokach}, \bibinfo{person}{Bracha Shapira}, {and} \bibinfo{person}{Paul~B. Kantor}} (Eds.). \bibinfo{publisher}{Springer}, \bibinfo{pages}{1--35}.
\newblock
\urldef\tempurl%
\url{https://doi.org/10.1007/978-0-387-85820-3\_1}
\showDOI{\tempurl}


\bibitem[Shehzad et~al\mbox{.}(2025)]%
        {IntentawareWorrying25}
\bibfield{author}{\bibinfo{person}{Faisal Shehzad}, \bibinfo{person}{Maurizio Ferrari~Dacrema}, {and} \bibinfo{person}{Dietmar Jannach}.} \bibinfo{year}{2025}\natexlab{}.
\newblock \showarticletitle{A Worrying Reproducibility Study of Intent-Aware Recommendation Models}. In \bibinfo{booktitle}{\emph{Proceedings of the 48th International ACM SIGIR Conference on Research and Development in Information Retrieval}} (Padua, Italy) \emph{(\bibinfo{series}{SIGIR '25})}. \bibinfo{publisher}{Association for Computing Machinery}, \bibinfo{address}{New York, NY, USA}, \bibinfo{pages}{3155–3164}.
\newblock
\showISBNx{9798400715921}
\urldef\tempurl%
\url{https://doi.org/10.1145/3726302.3730307}
\showDOI{\tempurl}


\bibitem[Sun(2023)]%
        {FreshSun23}
\bibfield{author}{\bibinfo{person}{Aixin Sun}.} \bibinfo{year}{2023}\natexlab{}.
\newblock \showarticletitle{Take a Fresh Look at Recommender Systems from an Evaluation Standpoint}. In \bibinfo{booktitle}{\emph{{ACM} {SIGIR}}}. \bibinfo{pages}{2629--2638}.
\newblock
\urldef\tempurl%
\url{https://doi.org/10.1145/3539618.3591931}
\showDOI{\tempurl}


\bibitem[Sun(2024)]%
        {SunBeyond24}
\bibfield{author}{\bibinfo{person}{Aixin Sun}.} \bibinfo{year}{2024}\natexlab{}.
\newblock \showarticletitle{Beyond Collaborative Filtering: {A} Relook at Task Formulation in Recommender Systems}.
\newblock \bibinfo{journal}{\emph{{SIGWEB} Newsl.}} \bibinfo{volume}{2024}, \bibinfo{number}{Spring} (\bibinfo{year}{2024}), \bibinfo{pages}{1--11}.
\newblock
\urldef\tempurl%
\url{https://doi.org/10.1145/3663752.3663756}
\showDOI{\tempurl}


\bibitem[Tsukuda and Goto(2020)]%
        {tsukuda2020explainable}
\bibfield{author}{\bibinfo{person}{Kosetsu Tsukuda} {and} \bibinfo{person}{Masataka Goto}.} \bibinfo{year}{2020}\natexlab{}.
\newblock \showarticletitle{Explainable recommendation for repeat consumption}. In \bibinfo{booktitle}{\emph{ACM RecSys}}. \bibinfo{pages}{462--467}.
\newblock


\bibitem[Wang et~al\mbox{.}(2019)]%
        {wang2019modeling}
\bibfield{author}{\bibinfo{person}{Chenyang Wang}, \bibinfo{person}{Min Zhang}, \bibinfo{person}{Weizhi Ma}, \bibinfo{person}{Yiqun Liu}, {and} \bibinfo{person}{Shaoping Ma}.} \bibinfo{year}{2019}\natexlab{}.
\newblock \showarticletitle{Modeling item-specific temporal dynamics of repeat consumption for recommender systems}. In \bibinfo{booktitle}{\emph{WWW}}. \bibinfo{pages}{1977--1987}.
\newblock


\bibitem[Zangerle and Bauer(2022)]%
        {EvaluationSurvey22}
\bibfield{author}{\bibinfo{person}{Eva Zangerle} {and} \bibinfo{person}{Christine Bauer}.} \bibinfo{year}{2022}\natexlab{}.
\newblock \showarticletitle{Evaluating Recommender Systems: Survey and Framework}.
\newblock \bibinfo{journal}{\emph{ACM Comput. Surv.}} \bibinfo{volume}{55}, \bibinfo{number}{8}, Article \bibinfo{articleno}{170} (\bibinfo{date}{Dec.} \bibinfo{year}{2022}), \bibinfo{numpages}{38}~pages.
\newblock
\showISSN{0360-0300}
\urldef\tempurl%
\url{https://doi.org/10.1145/3556536}
\showDOI{\tempurl}


\bibitem[Zou and Sun(2025)]%
        {zou2025survey}
\bibfield{author}{\bibinfo{person}{Kuan Zou} {and} \bibinfo{person}{Aixin Sun}.} \bibinfo{year}{2025}\natexlab{}.
\newblock \showarticletitle{A Survey of Real-World Recommender Systems: Challenges, Constraints, and Industrial Perspectives}.
\newblock \bibinfo{journal}{\emph{CoRR}}  \bibinfo{volume}{abs/2509.06002} (\bibinfo{year}{2025}).
\newblock
\urldef\tempurl%
\url{https://doi.org/10.48550/arXiv.2509.06002}
\showDOI{\tempurl}
\showeprint[arXiv]{2509.06002}


\bibitem[Zou et~al\mbox{.}(2024)]%
        {Hesitation24}
\bibfield{author}{\bibinfo{person}{Kuan Zou}, \bibinfo{person}{Aixin Sun}, \bibinfo{person}{Xuemeng Jiang}, \bibinfo{person}{Yitong Ji}, \bibinfo{person}{Hao Zhang}, \bibinfo{person}{Jing Wang}, {and} \bibinfo{person}{Ruijie Guo}.} \bibinfo{year}{2024}\natexlab{}.
\newblock \showarticletitle{Hesitation and Tolerance in Recommender Systems}.
\newblock \bibinfo{journal}{\emph{CoRR}}  \bibinfo{volume}{abs/2412.09950} (\bibinfo{year}{2024}).
\newblock
\urldef\tempurl%
\url{https://doi.org/10.48550/ARXIV.2412.09950}
\showDOI{\tempurl}
\showeprint[arXiv]{2412.09950}


\end{thebibliography}
%\balance

\end{document}